\documentclass[aps,prb,twocolumn,amsmath,notitlepage,floatfix,footinbib,superscriptaddress,showpacs, showkeys]{revtex4-1}
\usepackage{graphicx}
\usepackage{amssymb}
\usepackage{mathptmx}
\usepackage[utf8]{inputenc}
\usepackage[english]{babel}
\usepackage[usenames,dvipsnames]{color}
\usepackage{wrapfig}
\usepackage{longtable}
\usepackage{array}
\usepackage{multirow}
\usepackage{hyperref}
\usepackage{tikz}
\usepackage{scalefnt}
\usepackage{todonotes}
\usepackage{titletoc}
\usepackage[title,titletoc]{appendix}

\newcommand{\kay}{\mathbf{k}}
\newcommand{\cue}{\mathbf{q}}
\newcommand{\up}{\uparrow}
\newcommand{\down}{\downarrow}
\newcommand{\str}{^{*}}

\begin{document}
\title{Mechanisms of finite-temperature magnetism in the three-dimensional Hubbard model}  

\author{Daniel Hirschmeier}
\affiliation{Institut f\"ur Theoretische Physik, Universit\"at Hamburg, Jungiusstra\ss e 9, D-20355 Hamburg, Germany}
\author{Hartmut Hafermann}
\altaffiliation{Part of this work was conducted while the author was at Institut de Physique Th\'eorique (IPhT), CEA, CNRS, 91191 Gif-sur-Yvette, France.}
\affiliation{Mathematical and Algorithmic Sciences Lab, France Research Center, Huawei Technologies Co. Ltd., 92100 Boulogne-Billancourt, France}
\author{Emanuel Gull}
\affiliation{Department of Physics, University of Michigan, Ann Arbor, Michigan 48109, USA}
\author{Alexander~I.~Lichtenstein}
\affiliation{Institut f\"ur Theoretische Physik, Universit\"at Hamburg, Jungiusstra\ss e 9, D-20355 Hamburg, Germany}
\author{Andrey~E.~Antipov}\email{aantipov@umich.edu}
\affiliation{Department of Physics, University of Michigan, Ann Arbor, Michigan 48109, USA}

\date{\today}

\begin{abstract}
We examine the nature of the transition to the antiferromagnetically ordered state in the half-filled three-dimensional Hubbard model using the dual-fermion multiscale approach. Consistent with analytics, in the weak-coupling regime we find that spin-flip excitations across the Fermi surface are important, and that the strong coupling regime is described by Heisenberg physics. In the intermediate interaction, strong correlation regime we find aspects of both local and non-local correlations. We analyze the critical exponents of the transition in the strong coupling regime and find them to be consistent with Heisenberg physics down to an interaction of $U/t=10$.
\end{abstract}

\pacs{71.27.+a, 75.10.Jm, 71.10.Fd}

\maketitle

\section{Introduction}

The formation of complex states of matter in quantum systems is one of the central subjects of condensed matter physics. 
A particular interest is attached to strongly correlated fermionic systems, where an interplay between multiple physical phenomena leads to a rich set of phases and transitions between them, including Mott metal-insulator transitions, superconductivity, and magnetism \cite{Dagotto1994,Kotliar2006}. 
Understanding the microscopic mechanism that triggers a phase transition in these systems is a task of formidable complexity \cite{Troyer2005} and generally only approximate solutions of basic models can be obtained. 

The prototypical model for electronic correlations is the fermionic Hubbard model \cite{Hubbard:1963,Gutzwiller1963}. 
It is a benchmark model for a variety of theoretical methods \cite{LeBlanc2015a} and cold atom experiments \cite{Esslinger2010,Hart2014,Jotzu2015}. 
In three dimensions and for the particle-hole symmetric case, the model has a finite temperature transition to the antiferromagnetically ordered state at all values of interaction strength $U$ \cite{Hirsch1987}. 
The character of this transition varies substantially in different parts of the phase diagram \cite{Moriya1984,Pruschke2003} and remains under debate.  

At weak interactions, unlike in the lower dimensional case \cite{Moukouri2001}, the appearance of a magnetic order is attributed to thermal spin-flip excitations of electrons across the Fermi surface \cite{Moriya1984}. 
At strong electron repulsion, the emergence of magnetic order is attributed to a spin-spin interaction of local moments \cite{Cyrot1970, Anderson1978}.
Away from these limits, analytical results are not available and these mechanisms may be coexisting, competing, or replaced by different physics.

Describing the properties in this regime has been attempted by employing a set of analytic partial summation methods, including the unrestricted Hartree Fock approximation \cite{VanDongen1991}, the two-particle self-consistent approach \cite{Dare1996,Dare1999}, strong coupling expansions \cite{Logan1996,Szczech1995}, the spin-fluctuation approach \cite{Singh1998}, and the quantum rotor approach \cite{Zaleski2008}, which partially resolve the phase diagram. 
Numerical approaches to the problem include the coupled cluster expansion \cite{Pan1997} and lattice Monte-Carlo simulations \cite{Hirsch1987,Scalettar1989,Staudt2000,Campos2004,Paiva2011, Kozik2013} at system sizes up to $10^3$ sites.

Simulations using the dynamical mean-field theory (DMFT)  \cite{Metzner1989,Georges1992,Georges1996,Kotliar2006} interpolate between the small- and large-interaction regimes and describe the formation of local moments upon the increase in interaction \cite{Ulmke1997,Pruschke2003}. 
The method includes all local correlations \cite{Metzner1989,Metzner:1991} and captures the Fermi-surface nesting mechanism of the transition \cite{Ulmke1997} but does not describe nonlocal correlations and predicts phase transitions of mean-field character \cite{Byczuk2002a,Freericks2003}. 
Cluster extensions of DMFT \cite{Lichtenstein2000, KotliarSavrasov:2001, Maier2005, Kent2005} provide good estimates for the critical temperature \cite{Kent2005}, Green's functions \cite{FuchsGull:2011} and thermodynamics away from the transition \cite{Fuchs2011b}, but spatial fluctuations are short-ranged and criticality is not accessible \cite{Maier2005}. 

Recent development of diagrammatic multiscale methods allowed the extension of these results by incorporating nonlocal correlations  \cite{Toschi2007,Rubtsov2008, Slezak2009,Rohringer2013,Taranto2013}.
A treatment using the dynamical vertex approximation (D$\Gamma$A), assuming an anomalous dimension exponent $\eta = 0$, showed critical exponents close to the Heisenberg values \cite{Rohringer2011} and described the separability of static and dynamic correlations in the spectral properties of the model \cite{Schafer2015}.

In this paper we study the antiferromagnetic transition in the half-filled, three-dimensional ($3$D) Hubbard model over a wide range of interaction strengths $U$ by means of a different multiscale scheme, the dual-fermion (DF) approach \cite{Rubtsov2008, HafermannLi2009}.
This scheme incorporates DMFT as a correlated initial point of a perturbation  expansion, therefore taking the Fermi-surface nesting physics into account.
The antiferromagnetic nonlocal fluctuations are included through a diagrammatic resummation of a particle-hole ladder series. The method is referred to as the ladder DF approach \cite{HafermannLi2009}. 
It has previously been shown to correctly capture criticality in the similar Falicov-Kimball model in different dimensions \cite{Antipov2014} and together with D$\Gamma$A indicated an exponential reduction in $T_c$ according to the Mermin-Wagner theorem in two dimensions \cite{Otsuki2014,Schaefer2015b}.  

We study spin correlation functions, critical exponents, and spectral properties of the model. 
By comparison to DMFT results we describe the impact of nonlocal correlations and spin-exchange processes. 
We show that the magnetic properties of the model are characterized by the Fermi-surface nesting physics at $U/t \lesssim 4$, by a combination of two distinct mechanisms in the strongly correlated regime $4 \lesssim U/t \lesssim 10$, and by local moment Heisenberg physics at large  $U/t\gtrsim 10$, where $t$ is the hopping amplitude.

\section{Model and Method}

We consider the particle-hole symmetric Hubbard model on a 3D simple cubic lattice: 
\begin{equation}\label{eq:hubbard}
H=\sum_{\kay, \sigma} \varepsilon_{\kay} n_{\kay,\sigma} + \sum_i U (n_{i,\uparrow} - \frac{1}{2}) (n_{i,\downarrow} - \frac{1}{2}).
\end{equation}
Here lattice sites and lattice momenta are labeled $i$ and $\kay$ respectively, $\sigma=\up,\down$ denotes the spin projection, $\epsilon_\kay = -2t(\cos k_x + \cos k_y + \cos k_z)$ is the electronic dispersion at momentum $\kay$, and $U$ is the Coulomb repulsion between two fermions residing at the same site. 
We take the hopping $t = 1$ as the unit of energy. 

The DF approach was developed in Ref. \onlinecite{Rubtsov2008}. 
Here we briefly introduce the underlying idea. A detailed derivation of the formalism can be found in Ref. \onlinecite{Hafermann2012}. 
The DF approach is suited for the problem at hand, because it allows us to address the Fermi surface nesting physics, local-moment formation, and criticality on the same footing.
To achieve this, one starts from the DMFT solution of the model, which places a quantum impurity at each lattice site. New, so-called DF degrees of freedom are then introduced through a Hubbard-Stratonovich transformation that couples lattice and DFs locally. The latter interact via the vertex functions of the impurity model and mediate the coupling between the impurities, which is absent in DMFT.
After integrating out the lattice fermions, the resulting dual action is
\begin{equation}\label{eq:dual_action}
\tilde{S} =-\sum_{\omega,\kay,\sigma} f^{*}_{\omega,\kay,\sigma}\left[\tilde{G}^{(0)}_{\omega,\kay,\sigma}\right]^{-1}f_{\omega,\kay,\sigma} + \sum_i \tilde{U}[f^{*}_i,f_i],
\end{equation}
where $f$ labels the new fermionic degrees of freedom and $\omega$ labels Matsubara frequencies. The bare propagator $\tilde{G}^{(0)}_{\omega,\kay,\sigma} =G^{\mathrm{DMFT}}_{\omega, \kay,\sigma} - g_{\omega,\sigma}$ represents the momentum-dependent correction to the DMFT Green's function. By virtue of the self-consistency condition,  $g_{\omega,\sigma} = \frac{1}{N} \sum_{\kay} G^{\mathrm{DMFT}}_{\omega, \kay,\sigma}$, the local part of the DMFT Green's function is identical to the Green's function of the impurity problem. Here $N$ denotes the volume of the system.
The interaction $\tilde U$ is composed of all reducible, fully antisymmetric $n$-particle vertex functions of the impurity,
$\tilde U[f_i\str,f_i] \approx -\frac{1}{4}\gamma^{(4)}_{i,\alpha\beta\delta\gamma}f\str_{i,\alpha}f_{i,\beta}f\str_{i,\delta}f_{i,\gamma} + \hdots$, where the lowest-order term contains the two-particle vertex function, defined as
\begin{equation}\label{eq:gamma4}\begin{aligned}
\gamma^{(4)}_{1234}  & = g^{-1}_{1} g^{-1}_{3} \left[ \langle c_{1}c_{2}^\dagger c_{3}c_{4}^\dagger \rangle_{\text{imp}} - g_{1}g_{3} (\delta_{12}\delta_{34} - \delta_{14}\delta_{32}) \right] g_{2}^{-1} g_{4}^{-1}.
\end{aligned}
\end{equation}
Combined indices $1=\{\omega_1,\sigma_1\}$ are used in Eq. (\ref{eq:gamma4}) to shorten notation. 

The complete solution of problem (\ref{eq:dual_action}) is equivalent to the lattice problem, (\ref{eq:hubbard}), and therefore not tractable. The DF approach proceeds by constructing a low-order approximation in terms of the new variables, which corresponds to a summation of classes of diagrams in the original variables. The noninteracting (Gaussian) ensemble of DFs is equivalent to the full DMFT solution of the problem \cite{Rubtsov2009}.
Because we are interested in the description of the antiferromagnetic phase transition we restrict the expansion to ladder-type diagrams in the particle-hole channel. The dominance of these diagrams can be justified by a power counting argument in $1/d$ \cite{Otsuki2014}. In two dimensions, this set of diagrams introduces critical long-range fluctuations destroying the spurious order introduced through the mean field and leads to the exponential decay of the spin-spin correlations \cite{Otsuki2014}.
We note that this ``second'' approximation leads to an overestimation of the critical temperature for high interaction strengths but has been found to yield the correct critical behavior \cite{Antipov2014}. Other approximations are, in principle, possible; for instance, providing feedback to the DMFT hybridization function might lead to an improved estimate of the critical temperature \cite{Otsuki2014} and a consideration of corrections from higher-order vertex functions may yield more precise results \cite{Katanin2013}. Here, however, we keep the DMFT hybridization function unchanged, allowing us to study the impact of nonlocal correlations in a controllable way.

The ladder DF makes the assumption that three-particle and higher order terms in the interaction $\tilde{U}$ have a small effect. Numerical evidence of the validity of this assumption has been provided in Ref. \onlinecite{HafermannLi2009}.
The full fermionic frequency dependence of the impurity vertex $\gamma$ is retained, while restricting it to a single bosonic frequency, $\Omega = 0$. This approximation is similar to the construction of an effective functional for paramagnons in the same model \cite{Hertz1974}, which has been used to describe critical properties of quantum many-body systems \cite{Hertz1976}. 
We have checked that inclusion of more bosonic frequencies in the vertex $\gamma$ only leads to minor numerical corrections to the transition temperature, and does not affect critical properties (see Appendix \ref{app:static_nonstatic}). This supports the observation in Ref. \onlinecite{Schafer2015} on the static nature of nonlocal correlations in the model.

The DMFT impurity problem is solved numerically using the continuous-time quantum Monte Carlo hybridization expansion method \cite{Hafermann2013,GullMillis:2011,WernerComanac:2006}. The impurity vertex function is obtained on $160\times 160$ fermionic frequencies.
We sample the Brillouin zone on a grid of N=$16\times16\times16$ points and evaluate the local and $\kay$-dependent Green's functions $g_{\omega}$ and $G_{\omega, \kay}$ and the static spin susceptibility in reciprocal space $\langle S^z(q) S^z(-q) \rangle(\Omega = 0)$ and real space $\langle S^z_i S^z_j \rangle (\Omega = 0)$, from which the critical temperature and exponents are extracted. 
The critical slowing-down close to the transition temperature $T_c$ is overcome using an annealing scheme with a gradual decrease in temperature. 
Finite-size effects are eliminated by requiring that the correlation length $\xi$ does not reach $L/6$, where $L = 16$ is the linear system size in units of the lattice constant. 
A detailed description of the calculation procedure is provided in Appendix \ref{app:calc_proc}.

\section{Results}

\begin{figure}[ht]
\begin{center}
\includegraphics[width=\columnwidth]{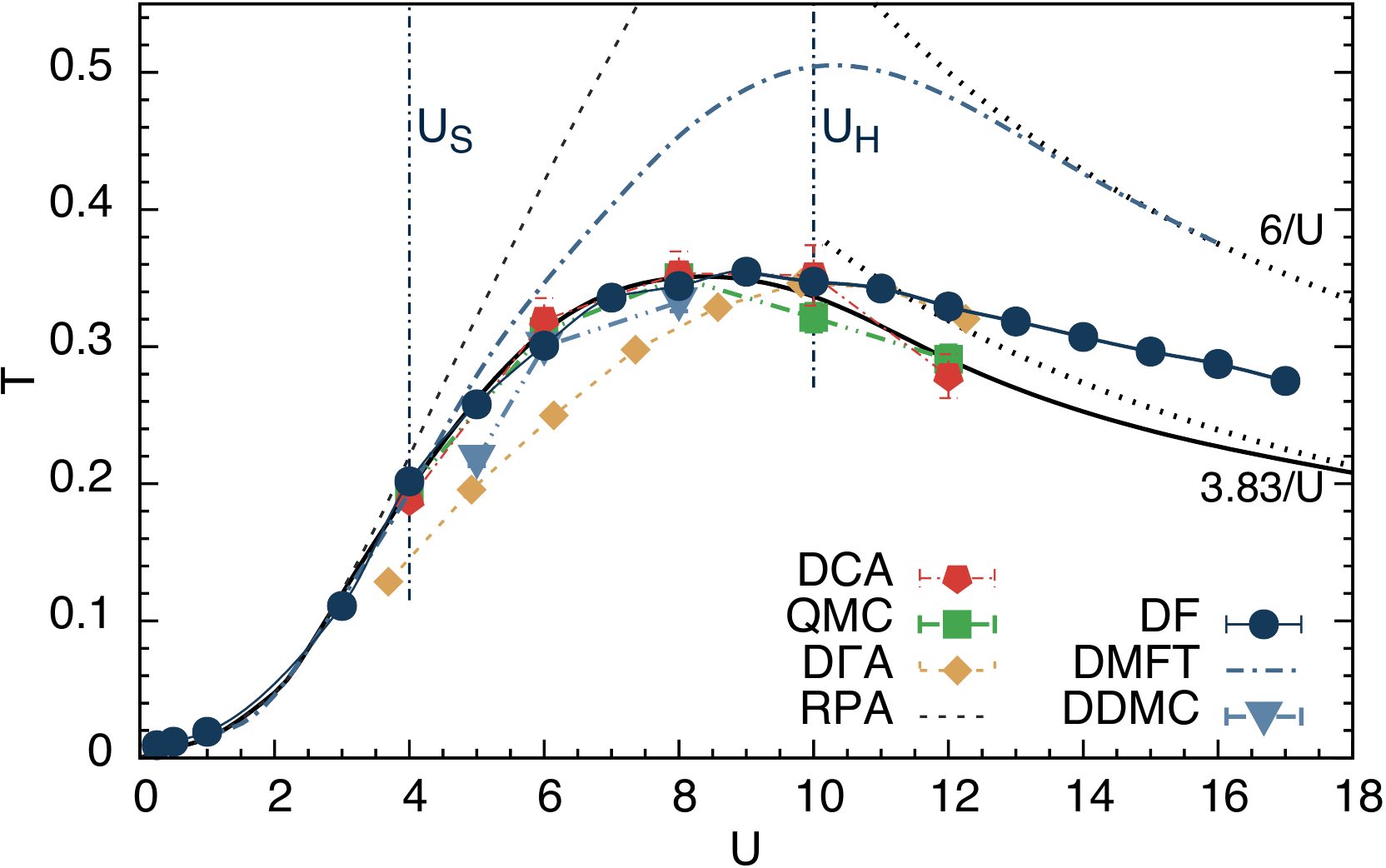}\end{center}\vspace*{-1.5em}
\caption{(Color online)Transition temperature $T_c$ as a function of $U$ (both in units of $t=1$) of the Hubbard model in three dimensions at particle-hole symmetry, as obtained by DF (filled dark circles) and  DMFT [dash-dotted (blue) line].
The comparison curves are obtained by the random phase approximation (RPA) from Refs. \onlinecite{Hirsch1987} and \cite{Scalettar1989} with a factor of 1/3 in $T_c$ to account for quantum fluctuations \cite{Freericks1994} (dashed black line), dynamical cluster approximation [DCA; filled (red) hexagons] \cite{Kent2005}, lattice quantum Monte Carlo [QMC; filled (green) squares]  \cite{Staudt2000}, determinantal diagrammatic Monte Carlo [DDMC; filled (blue) triangles] \cite{Kozik2013} and dynamical vertex approximation [D$\Gamma$A; filled (yellow) diamonds] \cite{Rohringer2011}. See also Ref. \onlinecite{Rohringer2011}. 
The combination of RPA values at small $U$ with the QMC \cite{Staudt2000}, the DCA \cite{Maier2005} and large-$U$ asymptotics is plotted as the thick solid black line.}
\label{fig:tc}
\end{figure}

Figure \ref{fig:tc} shows the phase boundary of the 3D Hubbard model obtained by a range of numerical methods. 
In the weak-coupling regime $T_c$ increases as a function of $U$; the slope is accurately described by the random-phase approximation (RPA; unrestricted Hartree-Fock) \cite{Hirsch1987} modified to account for quantum fluctuations up to $U \lesssim 4$ \cite{Freericks1994,MR1992}. 
In the strong-coupling regime an estimate is given by the Heisenberg limit $T_c = 3.83/U$, \cite{Sandvik1998, Affleck1988}, illustrated by the lower dotted line in Fig. \ref{fig:tc}. 
In the intermediate-coupling regime, estimates of $T_c$ are provided by the lattice quantum Monte Carlo \cite{Staudt2000} and dynamical cluster approximation (DCA) \cite{Kent2005}, extrapolated to the thermodynamic limit, which agree within their respective error bars and match the rescaled RPA value at $U=4$. 
Results of the determinantal diagrammatic Monte Carlo method \cite{Kozik2013} agree at $U=6$ and $8$ and provide a lower $T_c$ at $U\leq 5$. 
The results from the ladder D$\Gamma$A method \cite{Rohringer2011} are close, but nonlocal corrections seem slightly overestimated in the intermediate- to weak-coupling regime. 
In order to highlight the overall trend, a combined curve interpolated from RPA values at small $U$, averaged lattice quantum Monte Carlo \cite{Staudt2000} and DCA\cite{Maier2005} data, and large-$U$ asymptotics is shown in black.

The DMFT results, shown as dashed lines in Fig.\ref{fig:tc}, match the RPA values up to $U \approx 4$. 
Spatial correlations become important at $U > 4$, where DMFT overestimates the value of $T_c$.
At large $U$ DMFT approaches a limit of $6/U$, coinciding with the mean-field estimate \cite{Takahashi1977} (upper dotted line). 

This paper presents ladder DF results, shown in Fig. \ref{fig:tc} as dark solid circles. 
The difference between DF and DMFT results characterizes the contribution from nonlocal magnetic fluctuations. 
At $U < 4$ the DF method shows no deviation from the RPA/DMFT results, confirming the validity of the local approach.
In the intermediate-coupling regime $4 \lesssim U \lesssim 10$, nonlocal corrections are substantial and DF results for the transition temperature are consistent with lattice quantum Monte Carlo and DCA within error bars. 
As $U$ is increased further, an improvement on DMFT but deviation of $T_c$ to the best estimate and the high-temperature series expansion values becomes apparent. 
This is evidence of the relevance of diagrams that are not included.

We mark two important values of $U$ in Fig. \ref{fig:tc}. 
The point $U_\text{S} = 4$ marks the upper limit of validity of DMFT and hence the point where nonlocal correlations become important.  
Coincidentally this point is characterized by a change in concavity of the $T_c(U)$ curve, which does not occur in RPA calculations. 
We also add a point $U_\text{H} = 10$, at which the DF $T_c$ begins to deviate from the best available estimate.
This is explained in more detail in the following.

\begin{figure}[ht]
\begin{center}
\includegraphics[width=\columnwidth]{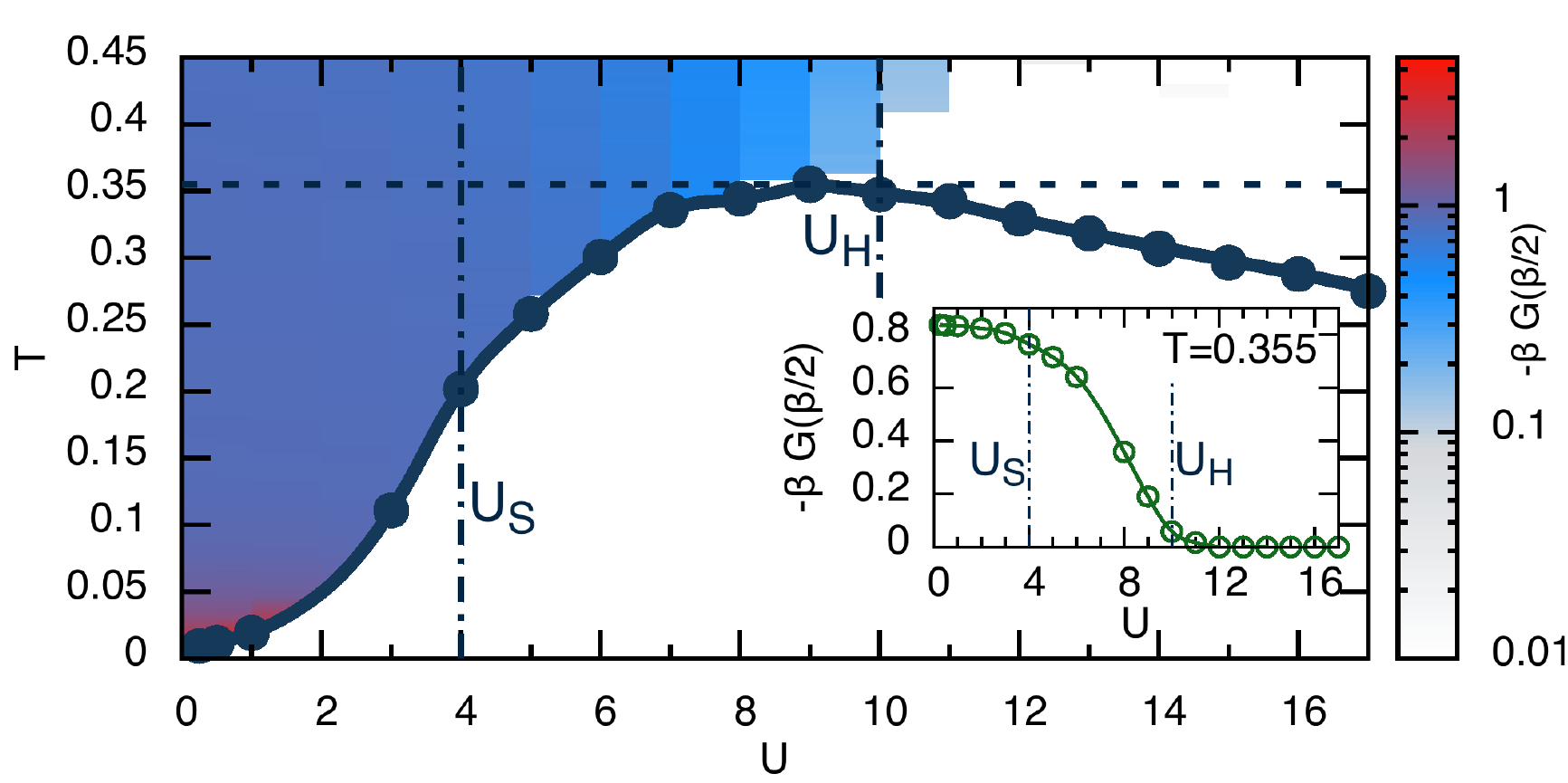}\end{center}\vspace*{-1.5em}
\caption{(Color online) Estimate for the density of states at the Fermi level, $-\beta G(\beta / 2)$, as a function of the temperature $T$ and interaction strength $U$. Inset: The cut at constant temperature $T=0.355$.}
\label{fig:dos}
\end{figure}

Figure \ref{fig:dos} shows an estimate of the density of states at the Fermi level $A(\omega = 0) \simeq -\beta G(\beta / 2)$. 
The solid line separates the isotropic phase from the symmetry broken antiferromagnetic region. 
A horizontal cut for $T = 0.355$ (dashed line) is shown in the inset, illustrating a metal-to-insulator crossover. 
For small $U\lesssim U_\text{S}$ the isotropic phase exhibits Fermi liquid metallic behavior.
As $U$ is increased the density of states decreases and shows pseudo-gap like behavior \cite{FuchsGull:2011} and at $U \simeq 10$ it becomes completely suppressed. 
We define $U_\text{H}$ as the point where $A(0) \rightarrow 0$ and a charge gap opens.

\begin{figure}[ht]
\begin{center}
\includegraphics[width=\columnwidth]{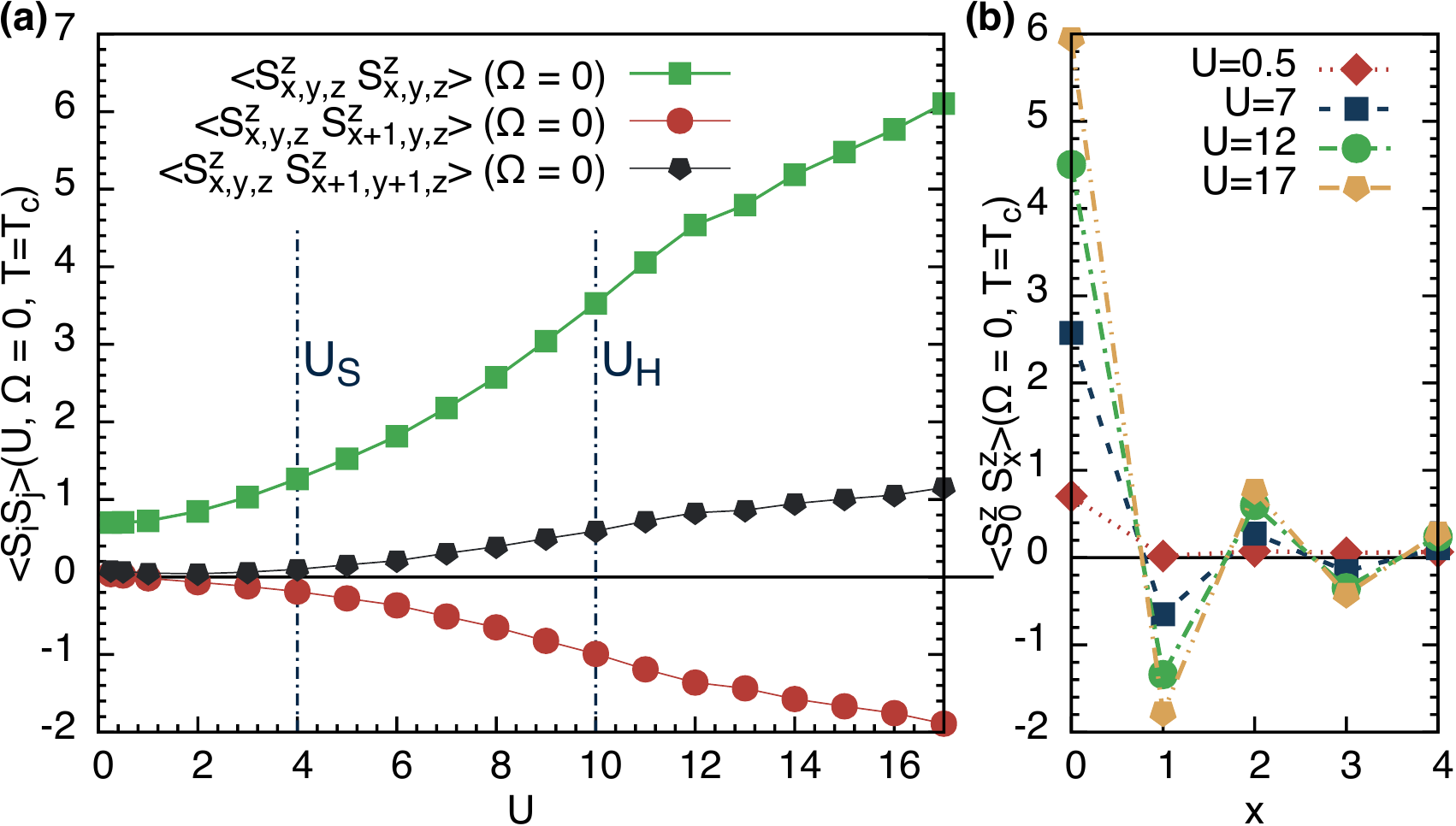}\end{center}\vspace*{-1.5em}
\caption{(Color online) (a) Local, nearest neighbor and next-nearest neighbor spin-spin correlation function as a function of interaction $U$ at $T \gtrsim T_c$. The formation of a well-defined magnetic moment occurs at large $U$. (b) Spatial dependence of the spin susceptibility along the real $x$ axis at different values of $U$ at $T \gtrsim  T_c$. }
\label{fig:sz_sz}
\end{figure}

Figure \ref{fig:sz_sz}(a) shows the magnitude of the local, nearest-neighbor, and diagonal spin-spin correlators $\langle S_z(i) S_z (j) \rangle (\Omega = 0)$ in  proximity to the transition approached from the high-temperature side, $T \gtrsim T_c$.
We observe that nonlocal correlations, represented by the spatial dependence of the spin-spin correlation function, are present for all values of $U$ and strongly increase for $U>U_\text{S}$, consistent with the deviation in $T_c$ between DMFT and DF shown in Fig. \ref{fig:tc}. 
Fig. \ref{fig:sz_sz}(b) shows the spatial extent of the spin-spin correlations for different values of $U$. It illustrates the antiferromagnetic nature of the model and shows that spin correlations are small for $U < U_\text{S}$.

\begin{figure}[t]
\begin{center}
\includegraphics[width=\columnwidth]{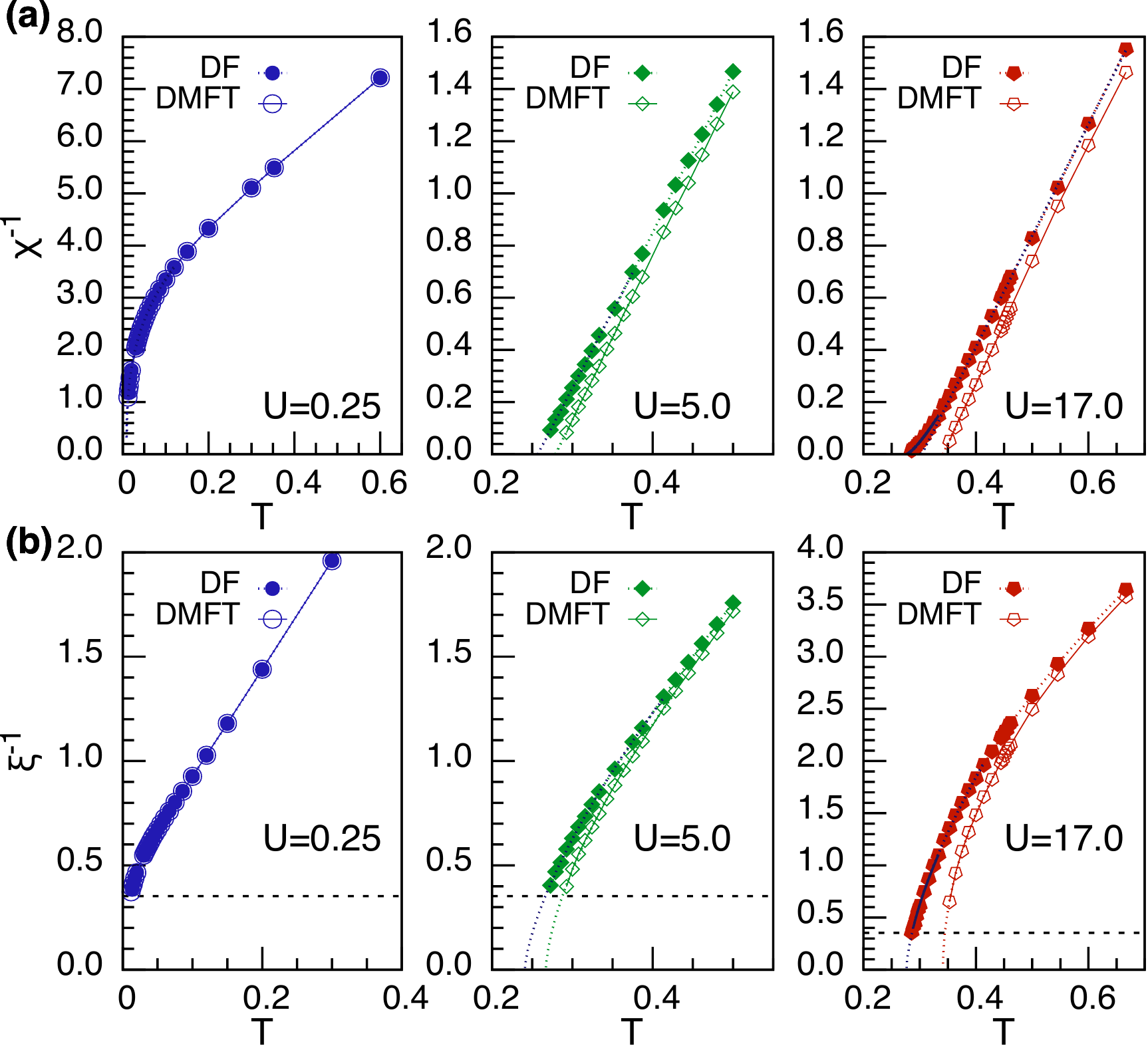}\end{center}\vspace*{-1.5em}
\caption{(Color online) (a) Inverse static magnetic susceptibility at $q=(\pi, \pi, \pi)$ and $U/t=0.25; 5.0; 17.0$ as obtained within the DF (filled symbols, dotted lines) and DMFT (open symbols, solid lines) methods and (b) corresponding correlation length. Dashed lines show the extrapolated low-temperature behavior, whereas dashed horizontal lines in (b) indicate the cutoff value for the inverse correlation length $\xi^{-1}_c=6/L$.}
\label{fig:df_dmft_susc}
\end{figure}

To characterize the influence of nonlocal correlations, we examine the temperature dependence of the susceptibilities and correlation lengths in Fig.~\ref{fig:df_dmft_susc}. 
Plotted is the temperature dependence of the static inverse magnetic susceptibility [Fig. \ref{fig:df_dmft_susc}(a)] and the inverse correlation length [Fig. \ref{fig:df_dmft_susc}(b)] at $q=(\pi, \pi, \pi)$ and $U = 0.25$ (left), $U = 5$ (center), and $U = 17$ (right) as obtained by the ladder DF method and DMFT at temperatures where the correlation length $\xi < L/6$. 
At small $U$ DMFT and DF coincide, as expected from Figs. \ref{fig:tc} and \ref{fig:sz_sz}. In the intermediate-$U$ regime, a difference in the transition temperature (as also depicted in Fig. \ref{fig:tc}) is visible. 
At large $U$, the deviation of DF from the DMFT data is substantial and a critical region with different temperature dependence is visible in both plots. 

\begin{figure}[t]
\begin{center}
\includegraphics[width=\columnwidth]{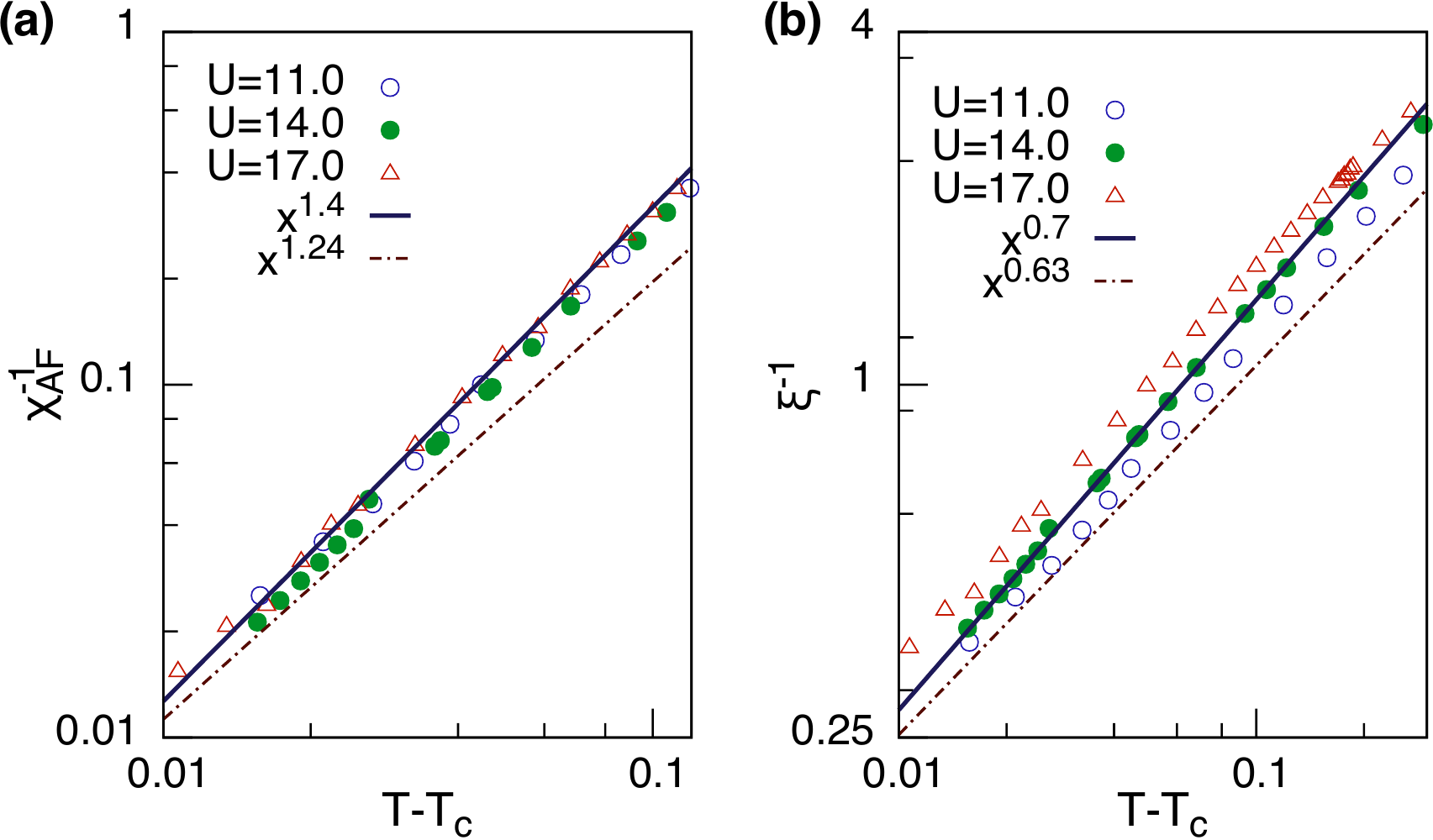}\end{center}\vspace*{-1.5em}
\caption{(Color line) (a) Inverse magnetic susceptibility and (b) inverse correlation length as a function of $T-T_c$ at $U/t=11$, $14$, and $17$, plotted on a logarithmic scale. 
The thick solid line represents the power-law dependence with Heisenberg exponent (a) $\gamma \sim 1.4$ and (b) $\nu = 0.7$ and the dashed line shows the power-law line with the Ising critical exponent (a) $\gamma \sim 1.24$ and (b) $\nu = 0.63$.}
\label{fig:susc_cor_log}
\end{figure}

We now turn to the criticality of the phase transition at large $U$. 
Figure \ref{fig:susc_cor_log}(a) shows the temperature dependence of the inverse magnetic susceptibility at $q=(\pi, \pi, \pi)$ and in the large-$U$ regime (shown here for $U/t = 11, 14, 17$) as a function of $T-T_c$, plotted on a log-log scale. 
The power-law dependence is resolved in $1.5$ decades and can be described by a Heisenberg exponent of $\gamma = 1.4 \pm 0.05$ \cite{Holm1993}. 
It is substantially different from the Ising value $\gamma_{\mathrm{Ising}} = 1.24$, plotted by the dashed line in the same graph. This clearly shows the qualitative difference from the mean-field criticality of DMFT \cite{Byczuk2002a}. 
Figure \ref{fig:susc_cor_log}(b) shows the temperature dependence of the inverse correlation length, plotted on a log-log scale. 
While available data for the fit in the critical region in this case are limited to a single decade, there is no uncertainty related to the determination of  $T_c$, as the latter is obtained from Fig. \ref{fig:susc_cor_log}(a). 
This substantially increases the quality of the fit, which reveals the correlation length exponent of the Heisenberg universality class $\gamma = 0.7 \pm 0.01$, which is again clearly distinguished from the Ising exponent $\gamma_{\mathrm{Ising}} = 0.63$. 

\begin{figure}[t]
\begin{center}
\includegraphics[width=\columnwidth]{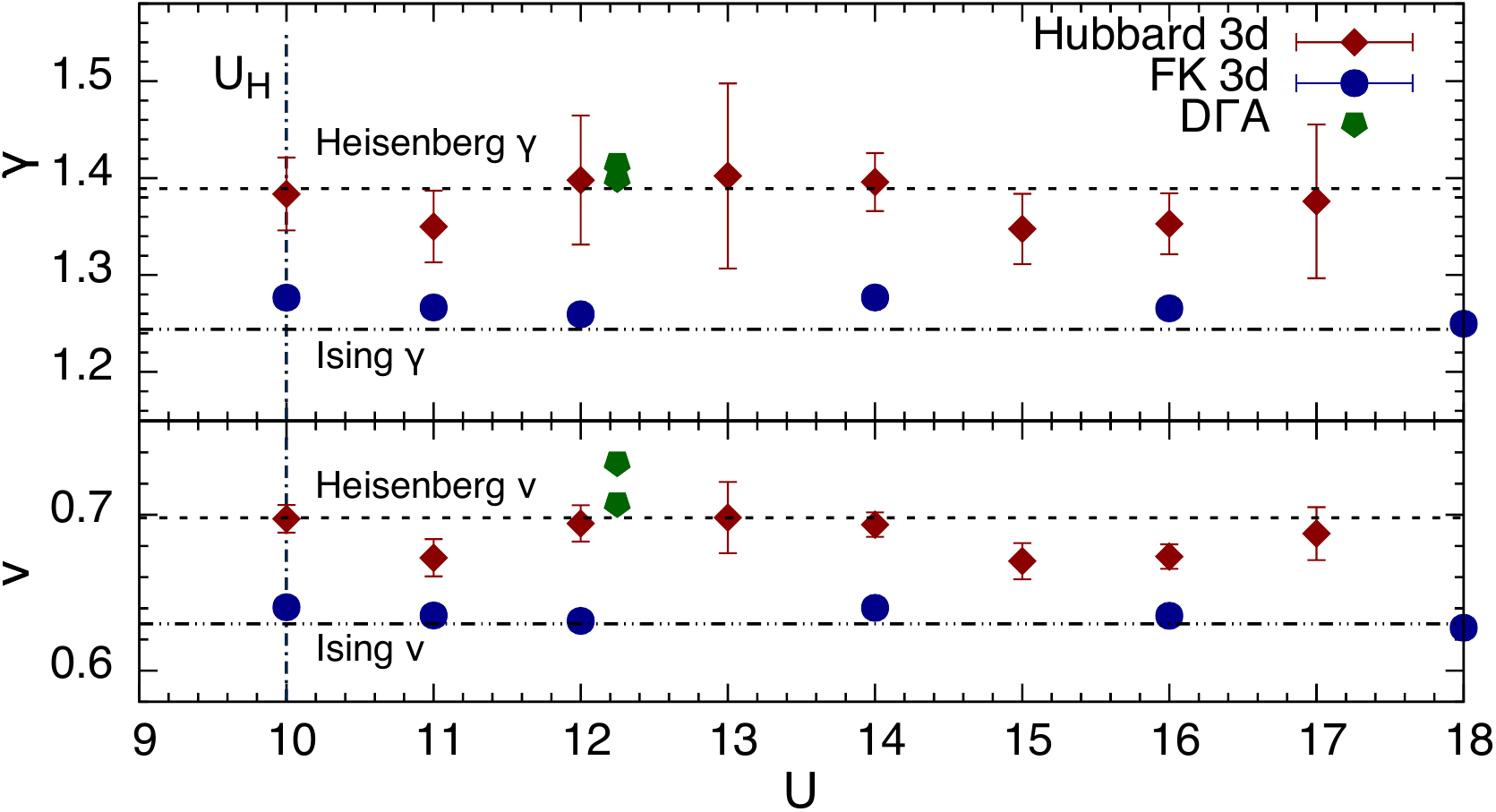}\end{center}\vspace*{-1.5em}
\caption{(Color line) Critical exponents $\gamma$ (susceptibility; top panel) and $\nu$ (correlation length; bottom panel) of the Hubbard model as obtained by the DF method at $U>U_H$ [filled (red) diamonds]. 
A comparison with the D$G$A result \cite{Rohringer2011} and the DF method for the Falicov-Kimball model \cite{Antipov2014} is provided. 
The Heisenberg and Ising exponents are shown as dotted and dash-dotted lines, respectively.}
\label{fig:exponents}
\end{figure}

A combined analysis of these results is presented in Fig. \ref{fig:exponents}. 
Shown are the values and fitting errors of the susceptibility exponent $\gamma$ (top) and the correlation length exponent $\nu$ (bottom) as a function of $U$ at $U \geq 10$. 
At lower values of $U$ the critical region with a different power-law divergence of the susceptibility and the correlation length, compared to DMFT, is not numerically accessible. 
The infinite-$U$ Heisenberg limit is also plotted, as the dashed line \cite{Holm1993}; the Ising result is the dot-dashed line. Results from the ladder D$\Gamma$A (two data sets) method \cite{Rohringer2011} are shown by (green) pentagons, DF results for the 3D Falicov-Kimball model \cite{Antipov2014}, obtained using the same fitting procedure, are plotted by filled (blue) circles, and our results for the 3D Hubbard model are depicted by (red) diamonds.
We find that, above an interaction strength of $U_\text{H} = 10$, our numerical estimates are consistent with the Heisenberg values. 

\section{Discussion}

Our results provide a straightforward picture of the magnetism of the half-filled 3D Hubbard model. 
At small values of $U < U_\text{S}$ the spin-spin interaction between local moments is not important, as indicated in Fig. \ref{fig:sz_sz}. 
The quantitative agreement between the exact transition temperature and the estimate, given by the DMFT and DF in Fig. \ref{fig:tc}, suggests that nonlocal corrections to the DMFT $T_c$ are small and hence the phase transition to the antiferromagnetic phase is well described by a local theory. 

The increase in $U$ leads to a gradual increase of the relevance of nonlocal fluctuations. 
This is not captured by the Fermi-surface nesting mechanism. 
In the intermediate-coupling regime $U_\text{S} \lesssim U \lesssim U_\text{H}$, $T_c$ continues to increase (indicative of a Fermi-surface nesting contribution to $T_c$) and nonlocal spin-spin correlations become non-negligible (indicative of Heisenberg physics), pointing to the coexistence of both scenarios, revealed in the decrease in the local density of states.
This shows some similarities to the $D\to\infty$ case, obtained by the DCA \cite{Gull2008c} and DMFT \cite{Taranto2012} methods, where the finite-temperature transition to the magnetic state is also present at all values of $U$ and is not destroyed by Goldstone bosons.
At $U = U_\text{H} \approx 10$ the density of states at the Fermi level vanishes, as shown in Fig. \ref{fig:dos}. 
Further increase in $U$ leads to a decrease in $T_c$ in accordance with the Heisenberg picture and in contradiction to the Fermi-surface nesting picture. 

Unlike the $T_c$, the critical exponents are universal and, as shown in Fig. \ref{fig:exponents}, are recovered by the DF method. 
The value of the exponents match the Heisenberg values precisely at $U \gtrsim U_H$, which confirms the validity of the Heisenberg picture in this parameter regime. 

At first glance it seems surprising that the DF method provides a correct set of exponents for different models and different dimensions (see also Ref. \onlinecite{Antipov2014} for an analysis of the Falicov-Kimball model), as, in general, an unbiased resummation of multiple competing channels is required and is normally done using a renormalization-group procedure. 
When only a single instability exists in the system a renormalization-group summation results in a ladder set of diagrams \cite{Giamarchi2003}. 
The fact that the correct values for the critical exponents are reproduced by the DF method indicates that only a single instability exists. 
This points to the importance of the correlated DMFT starting point, which includes all local correlations \cite{Zlatic1990,Metzner1989}. 
Validation of this conjecture by a renormalization-group procedure, formulated in the space of DFs \cite{Wentzell2015}, and comparison with more involved approximations beyond ladder diagrams, such as parquet diagrams \cite{Bickers1992,Bickers1991a,Jarrell2009,Jarrell2011}, is a direction for future research.

\section{Conclusions}
We have studied the antiferromagnetic transition in the particle-hole symmetric 3D Hubbard model by the ladder DF approach and quantified the relative importance of local and nonlocal correlations. 
We have characterized different regimes of the spin-ordering phase transition and inferred the physical mechanisms behind them.  
At $U \lesssim U_\text{S} \approx 4$, the physics is consistent with nesting of the Fermi surface. 
At $U_\text{S} \lesssim U \lesssim U_\text{H} \approx 10$, both Fermi-surface nesting and spin-ordering are important. 
This is the strongly correlated metallic regime, in which $T_c$ increases as a function of $U$. 
Spin fluctuations reduce the density of states on the Fermi level and the position of the critical temperature. However, the transition is still governed by the nesting of the Fermi surface.
As $U \gtrsim U_\text{H}$, a charge gap is present and nonlocal correlations become dominant. They manifest themselves in the spin-exchange mechanism of the ordering transition, resulting in a decrease in transition temperature upon the increase in $U$. 
The transition is described by Heisenberg critical exponents.

The ladder DF approach employed here provides a multiscale description, accounting for correlations on different length scales.
It may serve as a tool for studying phase transitions and criticality in strongly correlated systems, where the application of other numerical methods, such as the quantum Monte Carlo or the DCA, is difficult. 

\section{Acknowledgments} 

The authors thank Alexey N. Rubtsov, James P.~F. LeBlanc, Sergei~N.~Iskakov, Pedro Ribeiro, Olivier Parcollet and Mikhail I. Katsnelson for fruitful discussions. D.H. and A.I.L. acknowledge support from the DFG-FOR1346 program. H.H. acknowledges support from the FP7/ERC, under Grant Agreement No. 278472-MottMetals. A.E.A. was supported by DOE Grant No. ER 46932; E.G. was supported by the Simons Foundation. This work used the Extreme Science and Engineering Discovery Environment (XSEDE), which is supported National Science Foundation Grant No. ACI-1053575 and further computational resources provided by the HLRN-Cluster under Project No. hhp 00030.

\begin{appendix}
\renewcommand{\appendixname}{APPENDIX}

\setcounter{figure}{0}   \renewcommand{\thefigure}{A\arabic{figure}}
\setcounter{equation}{0} \renewcommand{\theequation}{A.\arabic{equation}}
\renewcommand{\thesubsection}{A.\Roman{section}.\Alph{subsection}}
\makeatletter
\renewcommand*{\p@subsection}{}  
\makeatother
\renewcommand{\thesubsubsection}{A.\Roman{section}.\Alph{subsection}-\arabic{subsubsection}}
\makeatletter
\renewcommand*{\p@subsubsection}{}  
\makeatother

\section{CALCULATION PROCEDURE}\label{app:calc_proc}
To obtain the building blocks for the dual-perturbation theory we solve the impurity problem numerically using the continuous-time quantum Monte Carlo hybridization expansion \cite{Hafermann2013}. 
We sample the single particle Green's function on 256 and the impurity vertex functions on $160$ fermionic and $1$ bosonic Matsubara frequency. 

The Brillouin zone is sampled on a grid of $N=16\times16\times16$ points.
The set of DF equations originates from a skeleton diagram expansion of the dual self-energy $\tilde{\Sigma}_{\omega,\kay}$. In the ladder DF approach, which is used in this work, all diagrams taken into account are of the particle-hole ladder type. Summing up all these ladder diagrams is most conveniently done by solving the Bethe-Salpeter equations for the charge and spin components of the renormalized particle-hole vertex defined by $\Gamma^{\text{ch/sp}}=\Gamma^{\up\up\up\up}\pm\Gamma^{\up\up\down\down}$ and then plugging this solution into the Schwinger-Dyson equation. The Bethe-Salpeter equations for the charge and spin components of the renormalized particle-hole vertex read
\begin{align}\label{eq:bse}
\Gamma^{\text{sp/ch}}_{\omega\omega'\Omega,\cue} &= \gamma^{\text{sp/ch}}_{\omega\omega'\Omega}+T\sum_{\omega''}\gamma^{\text{sp/ch}}_{\omega\omega''\Omega} \ \tilde{X}^{(0)}_{\omega''\Omega,\cue} \ \Gamma^{\text{sp/ch}}_{\omega''\omega'\Omega,\cue}\\
\tilde{X}^{(0)}_{\omega\Omega,\cue}&=\frac{1}{N}\sum_\kay \tilde{G}_{\omega,\kay}\tilde{G}_{\omega+\Omega,\kay+\cue} \label{eq:dual_ph_bubble} ,
\end{align}
where particle-hole notation is adopted and capital Greek letters are used to label bosonic Matsubara frequencies. Equation (\ref{eq:dual_ph_bubble}) defines the (dual) particle-hole bubble and $\gamma^{\text{sp/ch}}$ labels the spin and charge components of the impurity vertex. After solving Eq. (\ref{eq:bse}), either by inversion or iteratively, the renormalized vertex together with the Schwinger-Dyson equation yields the dual self-energy:
\begin{align}\label{eq:sde}
\tilde{\Sigma}_{\omega,\kay}=&-\frac{1}{2}T\sum_{\omega',\Omega,\cue}\gamma^{\text{ch}}_{\omega\omega'\Omega}\tilde{G}_{\omega'+\Omega,\kay+\cue}\tilde{X}_{\omega'\Omega,\cue}\left[\Gamma^{\text{ch}}_{\omega'\omega\Omega,\cue}-\frac{1}{2}\gamma^{\text{ch}}_{\omega'\omega\Omega,\cue} \right] \notag \\
&+\frac{3}{2}T\sum_{\omega',\Omega,\cue}\gamma^{\text{sp}}_{\omega\omega'\Omega}\tilde{G}_{\omega'+\Omega,\kay+\cue}\tilde{X}_{\omega'\Omega,\cue}\left[\Gamma^{\text{sp}}_{\omega'\omega\Omega,\cue}-\frac{1}{2}\gamma^{\text{sp}}_{\omega'\omega\Omega,\cue} \right].
\end{align}
Together with the Dyson equation for DFs $\left[\tilde{G}_{\omega,\kay}\right]^{-1}=\left[\tilde{G}^{(0)}_{\omega,\kay}\right]^{-1}-\tilde{\Sigma}_{\omega,\kay}$ Eqs. (\ref{eq:bse}) and (\ref{eq:sde}) form a set of nonlinear equations that can be solved self-consistently until convergence in $\tilde{G}_{\omega, \kay}$ is achieved, ensuring the approximation to be conserving in the Baym-Kadanoff sense, since all diagrams taken into account are dressed skeletons. The solution (fixed point) of the ladder DF equations is obtained by an iterative solution scheme up to a certain accuracy, which can be estimated by $\epsilon=\sum_{\omega,\kay}|\tilde{G}^{(n+1)}_{\omega,\kay}-\tilde{G}^{(n)}_{\omega,\kay}|$, where $\tilde{G}^{(n)}_{\omega,\kay}$ is the result for the Green's function obtained in the $n$th iteration step (here spin indices are omitted). 
A mixing between iterations as low as $\Xi=0.05$ is used because of the critical slowing-down at the phase transition.
Resulting dual Green's functions $\tilde{G}_{\omega,\kay}$ are in the convergence radius of the series summation if the eigenvalue problem,
\begin{equation}
\sum_{\kay,\omega'}\gamma_{\omega\omega'\Omega}\tilde{G}_{\omega',\kay}\tilde{G}_{\omega'+\Omega,\kay+\cue}\phi_{\omega'}=\lambda\phi_{\omega},
\end{equation}
yields a leading eigenvalue $\lambda_\text{max}<1$. In this case the underlying Bethe-Salpeter equation (BSE) can be solved by inversion. Note that due to the connection between the vertex and the susceptibility, a leading eigenvalue $\lambda_{max} = 1$ implies a divergent susceptibility. For leading eigenvalues below this threshold the spin susceptibility $\chi^{\text{sp}}$ is readily obtained from the spin component of the renormalized vertex $\Gamma^{\text{sp}}$ via
\begin{equation}\label{eq:spin_susc}
\chi^{\text{sp}}_{\Omega,\kay}=X^{(0)}_{\Omega,\kay}+\frac{T}{N}\sum_{\omega\omega'} X^{(0)}_{\omega\Omega,\kay}\Gamma^{\text{sp}}_{\omega\omega'\Omega} X^{(0)}_{\omega'\Omega,\kay}.
\end{equation}
Note that this relation holds for lattice fermions and dual fermions regardless, as long as the corresponding bubble $X^{(0)}$ or $\tilde{X}^{(0)}$ and vertex $\Gamma^{\text{sp}}$ or $\tilde{\Gamma}^{\text{sp}}$ are used. However, due to the correspondence between Green's functions of dual and lattice fermions at the single- and two-particle level, respectively (for details see Ref. \cite{bluebible}), it is possible to relate the spin susceptibility $\chi^{\text{sp}}$ for lattice fermions to the spin component of the renormalized \textit{dual} vertex $\tilde{\Gamma}^{\text{sp}}$ if Eq. (\ref{eq:spin_susc}) is rewritten as
\begin{align}
\chi^{\text{sp}} &=X^{(0)}+\bar{X}^{(0)}\ast\Gamma^{\text{sp}}\ast\bar{X}^{(0)}, \\
\bar{X}^{(0)} &=-\frac{1}{N} \sum_\kay G_{\omega,\kay}G_{\omega+\Omega,\kay+\cue}R_{\omega,\kay}R_{\omega+\Omega,\kay+\cue}.
\end{align}
In the equations above the asterisk is a shorthand notation to replace the convolution in Eq. (\ref{eq:spin_susc}) and the bubble has been replaced by $\bar{X}$ in the second term only. Hence the self-consistent solution of Eq. (\ref{eq:bse}) allows one to account for the influence of spatial correlations onto single and two-particle quantities.
 
Achieving convergence close to the phase transition is a delicate task, as one is confronted with the problem that the initial guess provided by $\tilde{G}^{(0)}_{\omega,\kay}$ is not within the convergence radius of the Bethe-Salpeter equation, while the solution is. One way to address this problem is to cut off the eigenvalues at 1 during the iterations until the Green's function lies within the convergence radius \cite{Otsuki2014}. Here we instead performed successive calculations at decreasing temperatures $T_0 > ... > T_l$, where we choose our starting point $T_0 > T_c^{\mathrm{DMFT}}$ such that the temperature is high enough for $\tilde{G}^{(0)}_{\omega,\kay}$ to be a sufficient initial guess. 
After obtaining a converged solution $\tilde{G}_{\omega,\kay}$ at temperature $T_m$, we interpolate this solution to the Matsubara frequency grid defined by $\omega_n=(2n+1)\pi T_{m+1}$ for every $\kay$ to obtain a suitable initial guess for the iterative solution of the DF equations at temperature $T_{m+1}$. The temperature is then lowered with every step, allowing for much faster convergence since the information about the spatial modulation of the self-energy is retained in the initial guess.

We perform temperature scans, as described above, at different fixed values of the Hubbard $U$ ranging from $0.25$ to $17$ in units of the hopping. 
The correlation length is obtained by fitting $\chi_{\text{fit}}^{-1}=c(1+2(k_x-\pi)^{2}\xi^{2})$ to the inverse spin susceptibility $[\chi^{\text{sp}}]^{-1}(\Omega=0,k_x,k_y=\pi,k_z=\pi)$ along the $k_x$-axis, with fit parameters~\cite{Mack2011} $c$ and $\xi$. 
We determine the critical exponents $\gamma$ and $\nu$ and the critical temperature $T_c$ by fitting a power law, $\chi^{-1}(T)=a(T-T_c)^{-\gamma}$, to the inverse spin susceptibility $\chi^{\text{sp}}(\Omega=0,\kay=\mathbf{R})^{-1}$ and the corresponding correlation length $\xi^{-1}(T)=b(T-T_c)^{-\nu}$ with fixed $T_c$ for $\xi$ respectively. 
The fits are performed within the so-called critical region, which is the temperature interval $\left[T_u,T_l\right]$ where the behavior of $\chi$ and $\xi$ deviates from the high-temperature mean-field dependence. 
While $T_l$ is given by the temperature at which the correlation length $\xi$ reaches $1/6$ of the linear system size $L$, it is more difficult to correctly determine the upper boundary $T_u$ of this critical region, which is crucial to avoid ambiguity in the fit. 
In order to do this we employ the procedure described in Ref. \onlinecite{Rohringer2011} and used in Ref. \onlinecite{Antipov2014} in the text and first determine the mean-field asymptotics of the inverse susceptibility by fitting a linear function, $\chi^{\text{lin}}$, to the high-temperature regime. 
The upper boundary of the critical region $T_u$ is determined as the temperature at which the ratio $r=\chi^{\text{sp}}/\chi^{\text{lin}}$ exceeds a certain value $r_c$. 
This way we can identify $T_u$ as the temperature where the spin susceptibility exits the linear mean-field regime. To eliminate the arbitrariness introduced by a specific choice of $r_c$, $T_u$ is determined for a set of values of $r_c=0.07 \div 0.15$. 

\begin{figure}[ht]
\begin{center}\includegraphics[width=\columnwidth]{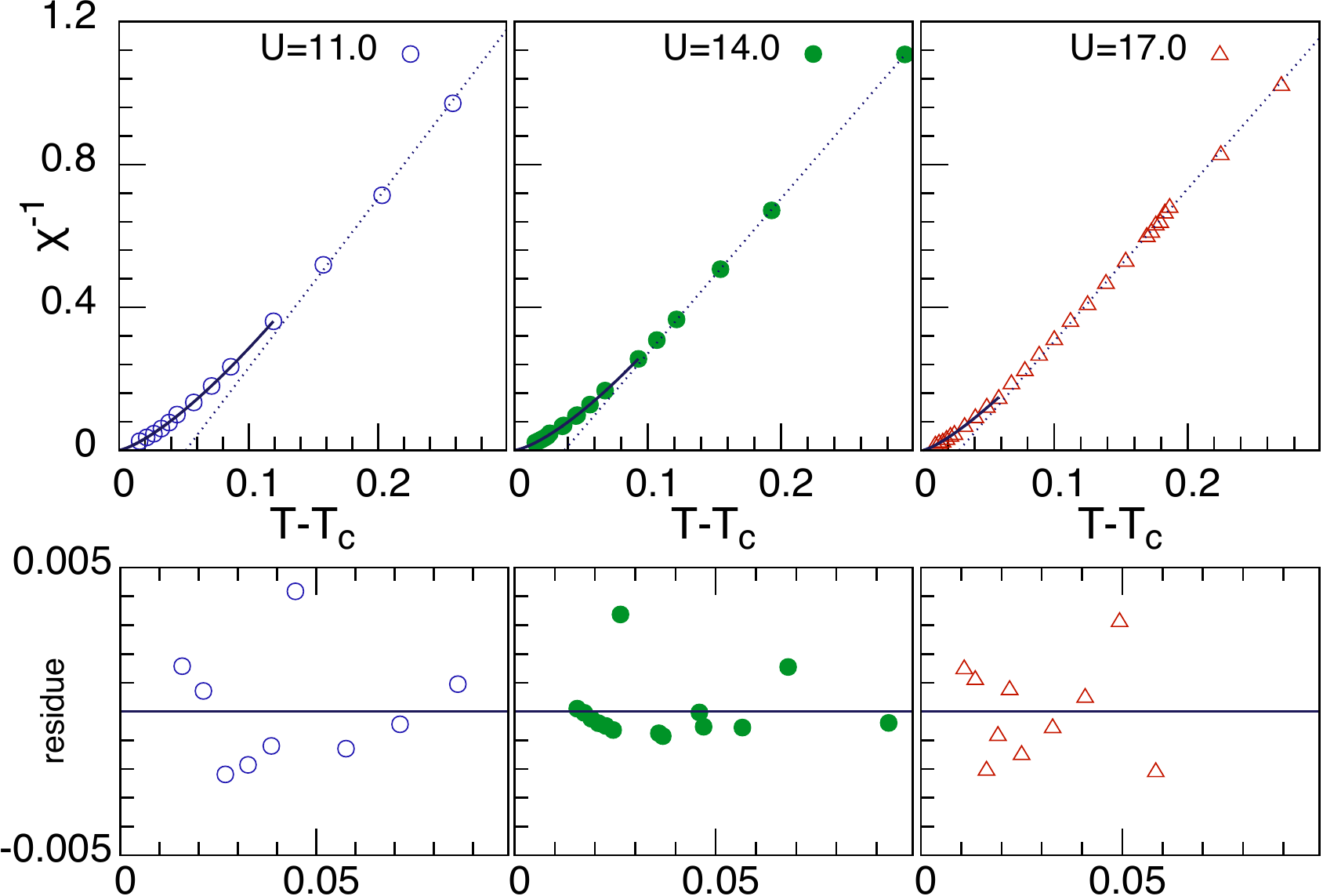}\end{center}\vspace*{-1.5em}
\caption{(Color online) Top: Inverse magnetic susceptibility at $\cue=(\pi,\pi,\pi)$ for $U/t=11, 14, 17$ as a function of $T-T_c$. Circles, DF values; dashed line, linear high-temperature mean-field fit; bold solid line, fit of critical exponent in the critical region. Bottom: Residue values of the susceptibility fit plotted versus the temperature.}
\label{fig:susc_fit}
\begin{center}\includegraphics[width=\columnwidth]{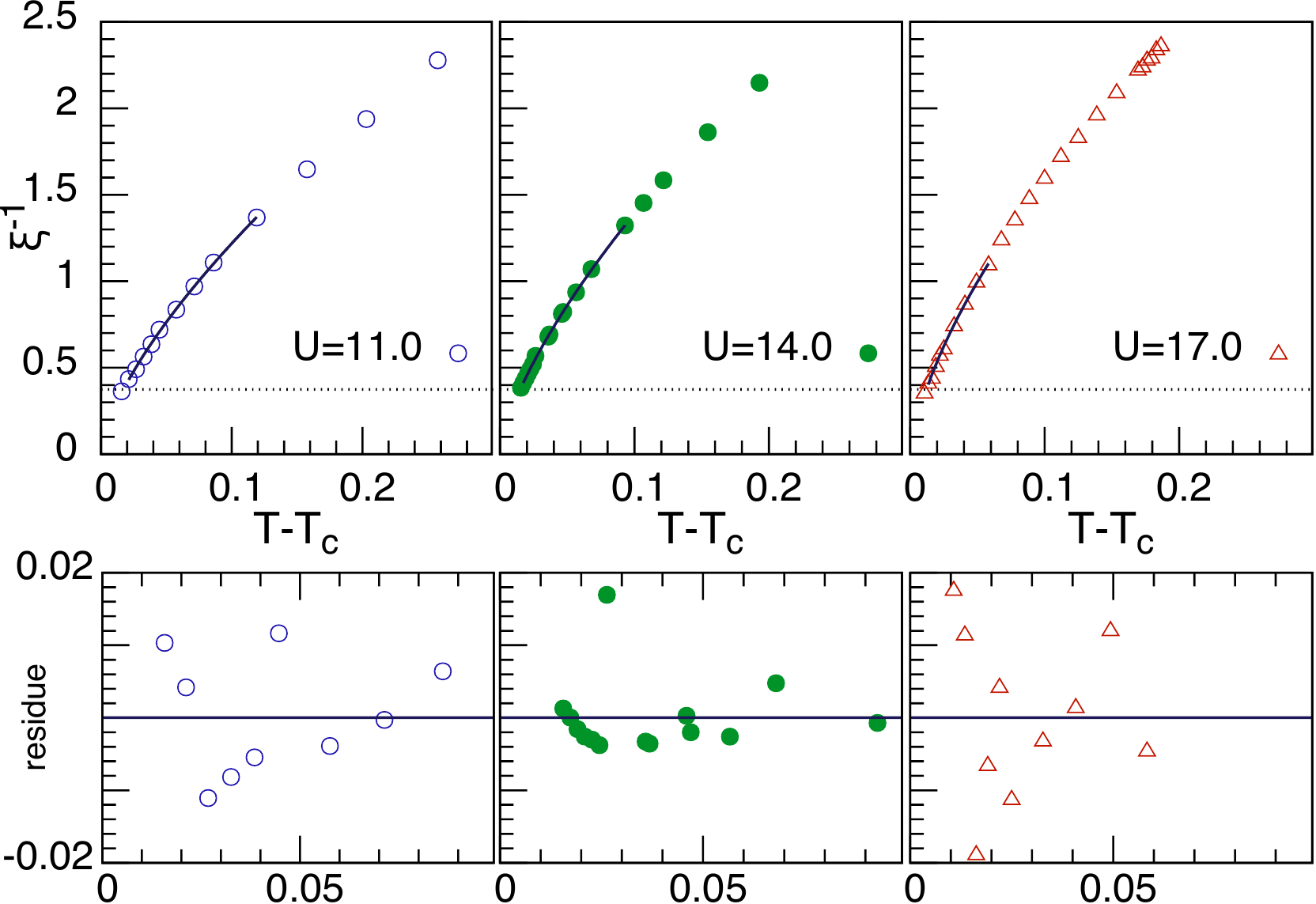}\end{center}\vspace*{-1.5em}
\caption{(Color online) Inverse correlation length (top) and residue values of the power-law fit (bottom) at $U/t=11, 14, 17$ as a function of $T-T_c$. The critical region, obtained from the data in Fig. \ref{fig:susc_fit}, is illustrated by the bold line.}
\label{fig:corr_fit}
\end{figure}

The final result for the critical exponents $\gamma, \nu$ and temperature $T_c$ is taken as the mean of the obtained fit parameters. The error bars are given by a quadrature sum of fit errors and averaging. 
As shown in the bottom panels in Figs. \ref{fig:susc_fit} and \ref{fig:corr_fit}, the quality of the fit is satisfactory, showing small temperature-independent residues.

\section{IMPACT OF FREQUENCY DEPENDENCE OF THE VERTEX}\label{app:static_nonstatic}

\begin{figure}[ht]
 \begin{center}
  \includegraphics[width=\columnwidth]{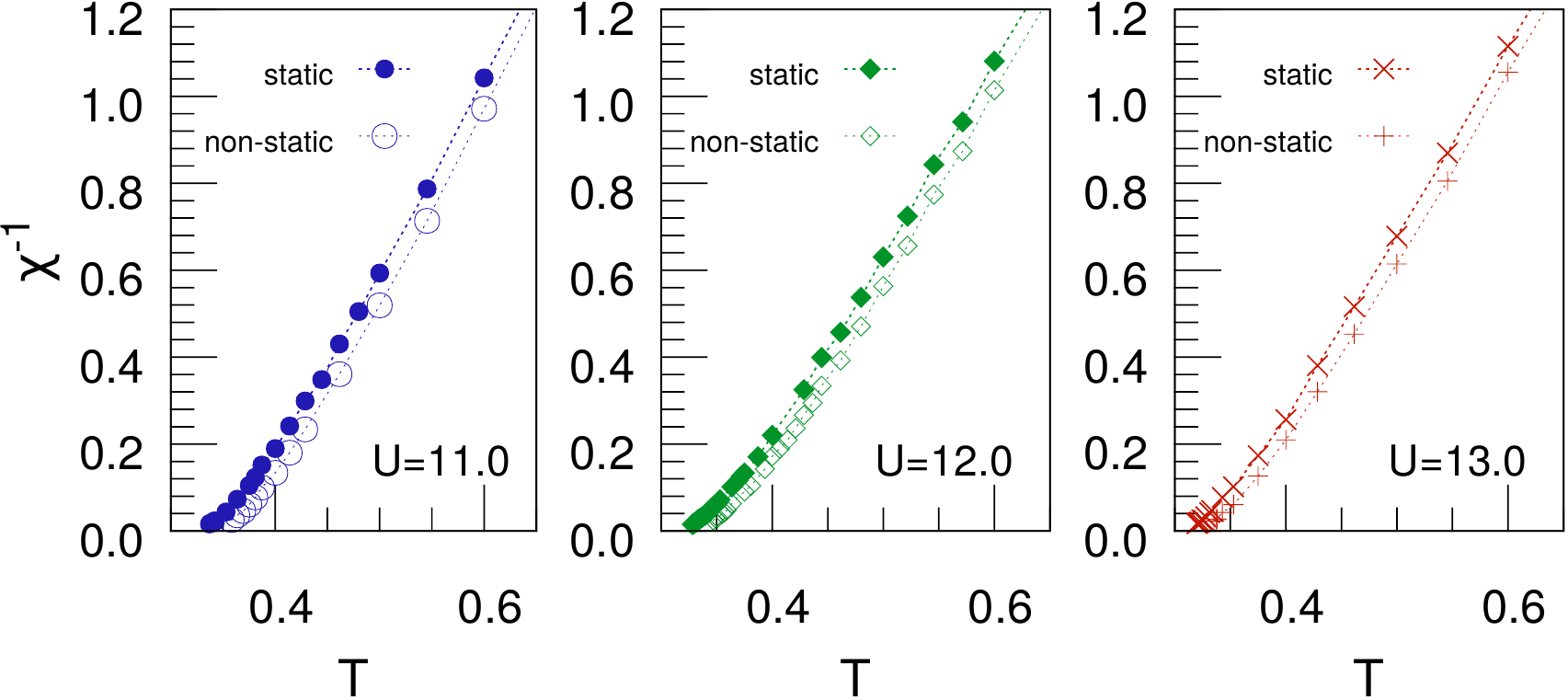}\vspace*{-1.5em}
  \caption{(Color online) Same as Fig. \ref{fig:susc_fit} for values of $U/t=11,12,13$ comparing the static approximation to the DF results at more than one bosonic frequency. Open symbols, static-approximation ladder DF approximation results; filled symbols, LDFA results using 120 bosonic frequencies. }
 \end{center}
 \label{fig:maize_and_blues}
\end{figure}

Figure \ref{fig:maize_and_blues} shows a comparison of the inverse magnetic susceptibilities obtained with the impurity vertex being sampled over 1 (static) or 120 (nonstatic) bosonic Matsubara frequencies, respectively. Two curves are separated by a constant offset, yielding a slightly lower value for the critical temperature in the nonstatic case. However both curves exhibit the same slope. Since we estimate the critical exponent $\gamma$ from a power-law fit to the inverse spin susceptibility $\chi^{-1}(T)=a(T-T_c)^{-\gamma}$, the static approximation will yield the same exponents as the nonstatic estimate since the constant offset is absorbed in the fit parameter $T_c$. Therefore the choice of the number of bosonic Matsubara frequencies does not affect the observed scaling quantities. Our approximation only takes low-energetic two-particle excitations into account, which seem to provide the leading divergent contributions to the static magnetic susceptibility and therefore correctly capture the criticality at the classical phase transition. In order to make a quantitative comparison of this approximation in a more general case, such as away from particle-hole symmetry, it is necessary to include more bosonic frequencies.

\end{appendix}
\bibliographystyle{apsrev4-1}
\bibliography{main}

\end{document}